\documentclass[]{JHEP3}%
\usepackage{graphics}
\usepackage{epsfig,latexsym}

\newcommand{\be}{\begin{equation}}
\newcommand{\ee}{\end{equation}}
\newcommand{\bea}{\begin{eqnarray}}
\newcommand{\eea}{\end{eqnarray}}
\newcommand{\lsim}{\mbox{\raisebox{-.6ex}{~$\stackrel{<}{\sim}$~}}}

\newcommand{\mx}{\mbox}

\newcommand{\p}{\partial}

\newcommand{\s}{\sigma}

\newcommand{\la}{\lambda}

\newcommand{\sH}{{\scriptscriptstyle H}}

\newcommand{\2}{\frac{1}{2}}

\newcommand{\LT}{\left[}
\newcommand{\RT}{\right]}
\newcommand{\LF}{\left(}
\newcommand{\RF}{\right)}
\newcommand{\ra}{\rightarrow}

\newcommand{\im}{\Longleftrightarrow}

\newcommand{\ie}{{\it i.e.}}

\newcommand{\cO}{{\cal O}}

\title{$p$-adic Inflation }
\author{Neil Barnaby, Tirthabir Biswas and James M.\ Cline\\
Physics Department, McGill University, Montr\'eal, Qu\'ebec, Canada H3A
2T8\\
E-mail: \email{barnaby@physics.mcgill.ca}, 
\email{tirtho@physics.mcgill.ca}, \email{jcline@physics.mcgill.ca}}

\preprint{}
\abstract{
We construct approximate
inflationary solutions rolling away from the unstable
maximum of $p$-adic string theory, a nonlocal theory with
derivatives of all orders.  Novel features include the existence of
slow-roll solutions even when the slow-roll parameters, as usually
defined, are much greater than unity, as well as the need for the Hubble
parameter to exceed the string mass scale $m_s$.
We show that the theory can
be compatible with CMB observations if $g_s / \sqrt{p} \sim
10^{-7}$, where $g_s$ is the string coupling, and if $m_s < 10^{-6}
M_p$. A red-tilted spectrum is 
predicted, and the scalar-to-tensor ratio is bounded from above as $r < 0.006$.  
The $p$-adic theory is shown
to have identical inflationary predictions to a local theory with superPlanckian
parameter values, but with the advantage that the $p$-adic theory 
is ultraviolet complete.
}
\begin{document}
\section{Introduction}
\label{sec:intro}

Many string theorists and cosmologists have turned their attention to building and
testing stringy models of inflation in recent years.  The goals have been to find
natural realizations of inflation within string theory, and novel features which would
help to distinguish the string-based models from their more conventional field theory 
counterparts.   The more popular categories include brane-antibrane \cite{brane}, D3/D7
\cite{D3D7}, modular \cite{modular}, DBI \cite{DBI} and tachyon-driven \cite{tachyon_inflation} 
inflation (see \cite{string_review} for a review).

In most examples to date, string theory has been used to derive an effective 4D field theory
operating at energies below the string scale.  Since string theory provides a complete
description of dynamics also at higher energies, it may be interesting to consider a model
which takes advantage of this distinctive feature.  This is usually daunting since the field
theory description should be supplemented by an infinite number of higher dimensional
operators at energies above the string scale, whose detailed form is not known.    In the
present work, we propose to take a small step in the direction of overcoming this barrier,
by considering a simplified model of string theory invented in 1987 \cite{witten}, in which
the world-sheet coordinates of the string are restricted to the field of $p$-adic numbers.  
Scattering amplitudes of open string theory can be related to those of the $p$-adic
strings.  A great advantage in  $p$-adic string theory is that it is possible to compute
{\it all} amplitudes of its lowest state and to determine a simple field-theoretic
Lagrangian which exactly reproduces them.  The result is a nonlocal field theory which is
nevertheless sensible in the far ultraviolet. 

The $p$-adic string resembles the bosonic string in that its ground state is a tachyon,
whose unstable maximum presumably indicates the presence of a decaying brane, analogous to
the unstable D25-brane of the open bosonic string theory \cite{bosonic}.  
Similarly to the bosonic string,
the potential is asymmetric around the maximum, with one direction leading to a zero-energy
vacuum, while in the other direction the potential is unbounded from below.  We will
consider whether it is possible to get successful inflation from rolling toward the bounded
direction.  This has been tried before in the context of the open string tachyon, and 
is difficult \cite{problems} because the tachyon potential is not flat enough to give a 
significant period of
inflation, and there are no parameters within the theory which can tune the potential to be
more flat.  In contrast, we will show that the $p$-adic string tachyon {\it can} roll slowly
enough to give many e-foldings of inflation.  There are two distinct regions of parameter
space which allow for successful inflation.  There is a region with $p = \mathcal{O}(1)$
for which the $p$-adic field potential is flat and slow roll inflation proceeds in the 
usual manner.  However, there is also a region of parameter space with $p \gg 1$ for which
the potential is \emph{extremely} steep ($|\eta| = M_p^2 |V''/ V|$ may be as large as $10^{11}$) 
but the 
$p$-adic scalar field nevertheless rolls slowly.  This remarkable behaviour relies on the 
nonlocal nature of the theory: the effect of the higher derivative terms in the action is 
to slow down the field sufficiently, despite its steep potential.  This new effect manifests 
itself only in the regime where the higher derivative interactions cannot be ignored.  It is 
also interesting that in our model the kinetic energy is responsible for driving inflation 
for a significant number of e-foldings, unlike in conventional models of inflation.

One may worry about the presence of these higher derivative terms, because they are 
usually known to introduce ghosts\footnote{One may also worry about classical
instabilities  which usually plague higher derivative theories; generically they go
by the name of  Ostrogradski instabilities (see \cite{woodard} for a review). These
are the  classical  manifestations  of having ghosts in the theory: they can have
arbitrarily large  negative energy, which leads to classical instability. Since the
nonlocal theory under consideration does not contain any ghosts we also do not
expect to find such instabilities. One way to see how such theories may avoid the
Ostrogradski instability  argument, valid for finite higher derivative
theories, is by noting  that one cannot construct the usual Ostrogradski Hamiltonian
because  there is no highest derivative in such nonlocal actions. 
Also, in arriving at the  Ostrogradski Hamiltonian, one assumes that all the 
derivatives of the field (except the maximal one) are independent canonical
variables. This is no longer true for theories with derivatives of infinite order. 
For instance, it is not possible to independently choose an  
initial condition with arbitrarily specified
values of all the  derivatives of the field. See ref.\
\cite{zwiebach} for a discussion of this point.} into the theory. In fact, it is easy
to check that for a scalar field theory  if one introduces only a  finite number of 
higher derivatives, then the model invariably  contains ghost degrees of freedom.
The reason why the $p$-adic string action can evade this  problem is  because it is
intrinsically nonperturbative in nature, the propagator being  modified in such a way
as to not contain any poles. In other words there are no physical  states, ghosts or
otherwise, around the true vacuum.  This novel way of  curing  the  problem of
ghosts in higher derivative theories,  while retaining some nice  properties such as
improved UV behaviour, was already  pointed out in \cite{warren} in the context of
gravity. It was also pointed out in  \cite{warren} that such theories can exibit
interesting new cosmological features. For instance,  one can obtain nonsingular
bouncing solutions by making gravity weak at short distances.  More recently, such
models have also been shown to possess inflationary solutions  \cite{justin}. 
However, the models of \cite{warren,justin} are phenomenological,  while the
$p$-adic action that we consider is an actual (albeit exotic) string theory
and  reproduces many nontrivial features of conventional string theories. 

We start by reviewing the salient features of $p$-adic string theory in section
\ref{sec:review}.  In section \ref{sect3} we show that this theory does not give
inflation if the higher-derivative terms in the action are ignored. However, near the top
of the potential, the energy can be large enough to justify keeping all higher derivative
terms.  In section \ref{solns} we show how to resum their contributions and we construct
approximate inflationary solutions valid near the top of the potential, by solving the
coupled equations for the tachyon and scale factor of the universe.  We give 
two different approximate methods in this section. 
In section \ref{fluct} we solve for the
fluctuations around this background to determine the power spectrum
of scalar and tensor perturbations which can be probed by the cosmic microwave background
(CMB).  There we show that it is possible to choose parameters which are compatible with
the measured amplitude and spectral index, and that the scalar-to-tensor ratio is bounded
from above as $r < 0.006$ in this model.  We also argue that it can be natural to
have initial conditions compatible with inflation in the $p$-adic theory.   We give
conclusions in section \ref{sec6}.  Appendix \ref{appa} gives details about the $p$-adic
stress tensor and the approximate inflationary solution of the Friedmann equation.  Appendix
\ref{appb} gives mathematical details about the incomplete cylindrical functions of the
Sonine-Schaefli form.  Appendix \ref{appc} explains a formal equivalence
between the dynamics of the $p$-adic tachyon and those of a local field theory with a
super-Planckian vacuum expectation value (VEV).

\section{Review of $p$-adic string theory}
\label{sec:review}
\begin{figure}[htbp]
\begin{center}
\includegraphics[width=0.45\textwidth,angle=0]{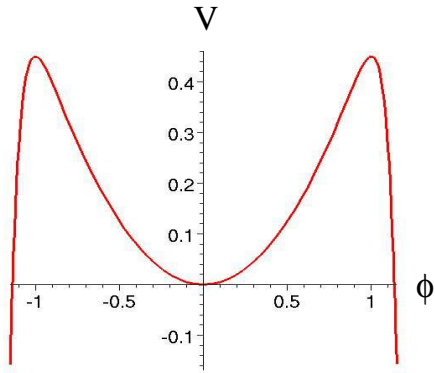}
\end{center}
\caption{The potential of the $p$-adic tachyon for $p=19$.\label{potfig}}
\end{figure}

The action of $p$-adic string theory is given by \cite{witten}
\begin{eqnarray}
S &=& 
{m_s^4\over g_p^2}\int d^{\,4}x\ \LF-\2\phi\, p^{-{\Box\over 2m_s^2}}\phi+{1\over p+1}\phi^{p+1}\RF
\nonumber \\
&\equiv& 
{m_s^4\over g_p^2}\int d^{\,4}x\ \LF-\2\phi\, e^{-{\Box\over m_p^2}}\phi+{1\over p+1}\phi^{p+1}\RF
\label{action}
\end{eqnarray}
where  $\Box = -\partial_t^2 + \nabla^2$ in flat space and we have defined
\be
{1\over g_p^2}\equiv {1\over g_s^2}{p^2\over p-1}\mx{ and } m_p^2\equiv {2m_s^2\over \ln p}
\label{gpmp}
\ee
The dimensionless scalar field $\phi(x)$ describes the open string tachyon, $m_s$ is
the string mass scale and $g_s$ is the open string coupling constant.  Though the action 
(\ref{action}) was originally derived for $p$ a prime
number, it appears that it can be continued to any postive integer and even makes sense
in the limit $p\rightarrow 1$ \cite{p=1}.  Setting $\Box=0$ in the action, the
resulting potential
takes the form $V = (m_s^4/g_p^2)(\frac12 \phi^2 - \frac{1}{p+1}\phi^{p+1})$.  Its shape
is shown in figure \ref{potfig}.

The action (\ref{action}) is a simplified model of the bosonic string which
only qualitatively reproduces some aspects of a more realistic theory.  
That being said, there are several nontrivial similarities between $p$-adic string theory and 
the full string theory.  For example, near the true vacuum of the theory $\phi=0$ the field 
naively has no particle-like excitations since its mass squared goes to 
infinity.\footnote{Reference \cite{zwiebach} found anharmonic oscillations around the vacuum 
by numerically solving the full nonlinear equation of motion. However, these 
solutions do not correspond to conventional physical
states.}\ \ This is the $p$-adic version of the statement
that there are no open string excitations at the tachyon vacuum.  A second similarity 
is the existence of lump-like soliton solutions representing $p$-adic D-branes 
\cite{Sen:$p$-adic}.  The theory of small fluctuations about these lump solutions has a spectrum 
of equally spaced masses squared for the modes \cite{Sen:$p$-adic},\cite{Minahan:ModeInteractions}, as in the case 
of normal bosonic string theory.

It should also be noted that the connection between (\ref{action}) and the DBI-type tachyon
actions, which have been widely studied in the literature in the context of tachyon matter
\cite{Sen:TachyonMatter}, is not entirely clear (see \cite{comparison} for a discussion
of the relation between $p$-adic and ordinary strings).  In the case of tachyon matter,
solutions which roll towards the vacuum $T \rightarrow \infty$ have late time asymptotics
$T \sim t$ and hence the tachyon never reaches this point \cite{DBIrolling}, whereas in the 
case of the $p$-adic string the vacuum is at a finite point in the field configuration space and 
homogeneous solutions rolling towards the vacuum typically pass this point without difficulty 
\cite{zwiebach} (at least in flat space).  In fact, the numerical studies of \cite{zwiebach} 
found \emph{no} homogeneous solutions which appeared to correspond to tachyon matter 
(vanishing pressure at late times).  This issue has been considered in the case of cubic
string field theory in \cite{taming}.  Related rolling tachyon solutions in cubic string
field theory have been discussed in \cite{CSFT}.

It is worth pointing out that one obtains very 
similar actions to (\ref{action}) with exponential kinetic operators 
(and usually assumed to have a cubic or quartic potential) when quantizing strings on random 
lattice \cite{random}. These  field theories are also known to reproduce several features, such 
as the Regge behaviour \cite{marc},  of their stringy duals. Although our analysis focuses on 
the specific $p$-adic action, it can easily be applied to such theories as well.  The connection
between $p$-adic string theory and ordinary string theory on a discrete lattice was explored
in \cite{comparison}.

The field equation that results from (\ref{action}) is
\be
\label{KG}
e^{-{\Box/m_p^2}}\,\phi=\phi^p
\ee
We are interested in perturbing around 
the solution $\phi=1$, which is a critical point of the potential,
representing the 
unstable tachyonic maximum. For odd $p$ one also has another unstable point at $\phi=-1$, 
but we will restrict our attention to solutions that start to evolve from the omnipresent 
$\phi=1$ maximum. In passing we also note that there is also the stable vacuum of the tachyon, 
at $\phi=0$.  For both even and odd $p$ the potential is unbounded from below.

It is also worth commenting on the physical interpretation of the fact that the potential is
unbounded from below.  The instability associated with the decay of the ``closed string
vacuum'' $\phi = 0$ to the ``true vacuum'' $\phi = \infty$ is thought to be associated
with the closed string tachyon instability \cite{bosonic}.

Notice that in the limit $p\rightarrow 1$ the equation of motion (\ref{KG}) becomes a local
equation
\[
  \Box \phi = 2 \phi \ln\phi
\]
Therefore $p\rightarrow 1$ is the limit of local field theory.  In the 
present work we will also consider a very different limit, $p \gg 1$, in which the nonlocal
structure of (\ref{action}) is playing an important role in the dynamics.

\section{Absence of Naive Slow Roll Dynamics}
\label{sect3}

One may wonder whether the field theory (\ref{action}) naively allows for slow roll inflation 
in the conventional sense.  Naively one might expect that for a slowly rolling field the higher
powers of $\Box$ in the kinetic term are irrelevant and one may approximate (\ref{action})
by a local field theory.  The action (\ref{action}) can be rewritten as
\begin{equation}
\label{local}
S = \int d^4 x \left[ \frac{1}{2}\chi \Box \chi - V(\chi) + \cdots \right]
\end{equation}
where we have defined the field $\chi$ as
\begin{eqnarray}
  \chi &=& \chi_0 \, \phi \label{canonical} \\
  \chi_0 &=& \frac{p}{g_s}\sqrt{\frac{\ln p}{2(p-1)}}\, m_s \label{chi0}
\end{eqnarray}
and the potential is
\begin{equation}
\label{pot}
  V(\chi) = \frac{m_s^2}{\ln p}\chi^2 
  - \frac{m_s^4}{g_s^2}\frac{p^2}{p^2-1}\left(\frac{\chi}{\chi_0}\right)^{p+1}
\end{equation}
In (\ref{local}) the $\cdots$ denotes terms with higher powers of $\Box$.  The truncation
(\ref{local}) is quite analogous to what is usually performed in the literature when one
studies inflation from the string theory tachyon \cite{tachyon_inflation}.  In the case of
the usual string theory tachyon corrections involving $\Box^2$ and higher are expected,
but the full infinite series of higher derivative terms is not known explicitly
(see, however, \cite{laidlaw} for a calculation of the string theory tachyon action
up to order $\partial^6$).  Thus, the standard approach is to simply neglect such terms.
We will show, however, that under certain circumstances the higher derivative corrections
may play an extremely important role in the inflationary dynamics.

Working in the context of the action (\ref{local}) let us consider the slow roll 
parameters describing the flatness of the potential (\ref{pot})
about the unstable maximum $\chi = \chi_0$.  It is straightforward to show that
\begin{eqnarray}
  \left. \frac{M_p^2}{2}\frac{1}{V(\chi_0)^2}
  \left(\frac{\partial V(\chi)}{\partial \chi}\right)^2\right|_{\chi=\chi_0}
  &=& 0 \label{V_slope}\\
  \left. M_p^2 \frac{1}{V(\chi_0)}
  \frac{\partial^2 V(\chi)}{\partial \chi^2}\right|_{\chi=\chi_0} &=& 
  -\frac{4 g_s^2}{\ln p}\frac{p^2-1}{p^2}\left(\frac{M_p}{m_s}\right)^2 \label{V_curv}
\end{eqnarray}
Thus one would naively expect that inflation is only possible in the theory (\ref{action}) 
for $g_s M_p /( \sqrt{\ln p}\, m_s)$ $\ll 1$.  For $p = \mathcal{O}(1)$ we will see that
this expectation is correct, however, for $p \gg 1$ this intuition  is incorrect and that 
successful inflation can occur even with 
$M_p^2 V^{-1} \partial^2 V/ \partial \chi^2 \sim -10^{16}\,$!   
The reason for this surprising result is that the dynamics of the $p$-adic tachyon field is set 
by the mass scale which 
appears in the kinetic term, $m_s$, rather than the mass scale which is naively implied by the 
potential 
\begin{equation}
\label{mass}
  m_\chi^2(\chi_0) \equiv 
  \left. \frac{\partial ^2 V}{\partial \chi^2} \right|_{\chi = \chi_0} 
  = -\frac{2(p-1)}{\ln p} m_s^2
\end{equation}
Clearly for $p \gg 1$ we have $|m_\chi^2| \gg m_s^2$.

\section{Approximate Solutions: Analytical techniques}
\label{solns}

In this section we construct the approximate solutions for the scalar field and
the quasi-de Sitter expansion of the universe,
in which $\phi$ starts near the unstable maximum ($\phi = 1$) of its potential
and rolls slowly toward the minimum ($\phi = 0$).  Cosmological solutions of 
similar nonlocal theories have also been considered in \cite{cosmo}. 

We use two different formalisms to construct inflationary solutions.  We first
devise a perturbative expansion in  $e^{\la t}$ similar to what was carried out in 
\cite{zwiebach} to study rolling solutions in  flat spacetime.  Our second formalism
is the analogue of the usual slow roll approximation: we assume that the friction term
in the $\Box$ operator dominates over the acceleration term and also neglect the time
variation of $H$.

We first discuss the perturbative expansion in powers of $e^{\lambda t}$.  
Our starting point is the ansatz 
\be
\label{phi_series}
\phi(t)=1-\sum_{r=1}^{\infty}\phi_re^{r\la t}
\ee
and
\be
\label{H_series}
H(t)=H_0-\sum_{r=1}^{\infty}H_re^{r\la t}
\ee
We have chosen the parameterisation such that at $t\ra -\infty$, $\phi$ starts from the top
of  the hill where the universe is undergoing de Sitter expansion with Hubble constant
$H_0$.  
As $t\ra -\infty$, $e^{\la t}\ra 0$ and all the correction terms, which quantify the
departure from the pure de Sitter phase,  vanish. As $t$
increases, the field rolls  toward the true vacuum $\phi = 0$, in fact reaching it at a
finite time. Classically, the model admits infinitely many e-foldings of inflation, 
although only the last 60 e-foldings before the end of inflation are relevant
for observation.  This idealized behaviour is an artifact of neglecting quantum
fluctuations; quantum mechanically the field cannot sit at $\phi = 1$ for an infinite
amount of time.  We will return to this issue later and show that the inclusion of quantum
fluctuations does not spoil inflation, which it would in ordinary local field theory if
the $\eta$ parameter is large.

Given the ansatz (\ref{phi_series},\ref{H_series}), we can expand the field equation for the
$p$-adic scalar (\ref{KG}) and the Friedmann equation as a series in $e^{\la t}$ and then
determine the coefficients $\{\phi_r,H_r\}$  systematically, order by order. We will show
that this can be done consistently. 
The zeroth order Klein-Gordon equation is trivially satisfied,  by virtue of the fact that 
we start from a  maximum of the potential.

\subsection{$p$-adic Scalar Field Evolution}

Let us first find an approximate solution for the scalar field equation of motion
(\ref{KG}).
We note that to compute quantities such as $\Box^n e^{\la t}$  to the order of
interest, it  is sufficient to use a truncation of (\ref{phi_series}) and (\ref{H_series}):
\be
\label{2nd_order_expasions}
\phi=1-u-\phi_2u^2,\qquad H=H_0-H_1u-H_2u^2
\ee
where we have used the freedom to choose the origin of time to set
$\phi_1 \equiv 1$, and for convenience we have defined the new variable
\be
\label{u}
u\equiv e^{\la t}
\ee
in terms of which the $\Box$ operator takes the form
\be
\Box=-\la^2\LT u^2{\p^2 \over \p u^2}+\LF1+{3H\over \la}\RF u\,{\p \over \p u}\RT
\ee
We wish to compute quantities such a $\Box^n e^{\la t}$ up to $\cO(u^2)$.
This can be done recursively. Writing $\Box^n\phi$ (where $n\ge 1$) as
\begin{equation}
\label{box_n_phi}
(-\Box)^n \phi =A_n u+B_n u^2+\dots
\end{equation}
and applying another $\Box$ operator one finds the following  recursion relations for the 
coefficients $A_n$ and $B_n$:
\be
\label{A_n+1}
A_{n+1}=A_n(\la^2+3H_0\la)
\ee
and 
\be
\label{B_n+1}
B_{n+1}=B_n(4\la^2+6H_0\la)-3H_1\la A_n
\ee
Equation (\ref{A_n+1}) has the solution
\be
A_n=a_1\left(\lambda^2 + 3H_0\lambda\right)^n
\ee
with $a_1 = -1$ (from examination of $\Box\phi$)
while a suitable ansatz for $B_n$ is given by
\be
B_n=b_1(\la^2+3H_0\la)^n+b_2(4\la^2+6H_0\la)^n
\ee
The coefficients $b_1,b_2$ can be deduced from (\ref{B_n+1}) using the
initial values
\be
b_1=-{H_1\over H_0+\la}\mx{ and }\ b_2={H_1\over H_0+\la}-\phi_2
\ee
which follow from explicitly computing $\Box\phi$ and $\Box^2\phi$.
Putting everything together we now have
\begin{eqnarray}
(-\Box)^n \phi &=& \delta_{n,0}-(\la^2+3H_0\la)^nu+\left[\LF {H_1\over H_0+\la}-\phi_2\RF(4\la^2+6H_0\la)^n
  \right. \nonumber \\
&& \left. -{H_1\over H_0+\la}(\la^2+3H_0\la) ^n\right] u^2 + \mathcal{O}(u^3)
\label{box_n_phi_soln}
\end{eqnarray}
which works also for the case $n=0$.
Using (\ref{box_n_phi_soln})  one can resum the contributions coming 
from all the powers of $\Box$ in the exponential operator $e^{-\Box / m_p^2}$ to give
\be
e^{-{\Box/m^2_p}}\,\phi=1-e^{\mu_1}u+\LT\LF\s-\phi_2\RF e^{\mu_2}- \s e^{\mu_1}\RT u^2 
+ \cdots
\label{eboxphi}
\ee
where we have introduced
\be
\mu_1\equiv {\la^2+3H_0\la\over m_p^2}\ \mx{ ,  }\ \mu_2\equiv {4\la^2+6H_0\la\over m_p^2}\ \mx{ and }\ \s\equiv {H_1\over H_0+\la} \label{mu_defs}
\ee
We conjecture that such resummations are possible for higher order terms as well. 
Notice that (\ref{eboxphi}) reduces to $\phi$ in the limit $m_p\to\infty$, as it should.

To solve the equation of motion for the scalar field, we must equate (\ref{eboxphi})
to the right-hand-side of (\ref{KG}):
\be
\phi^p=[1-u-\phi_2u^2]^p= 1-pu-p\LF\phi_2-{p-1\over 2}\RF u^2 + \cdots
\ee
matching coefficients for each order in $u$.
The zeroth order equation is identically satisfied, as promised earlier, while matching 
at first order gives 
\be
e^{\mu_1}=e^{(\la^2+3H_0)/ m^2_p}=p
\label{la}
\ee
which, using (\ref{gpmp}) can be rewritten in the form
\begin{equation}
\label{la_simple}
  \lambda^2 + 3H_0 \lambda = 2 m_s^2
\end{equation}
which is independent of $p$.  Later we will see that $H_0\gg m_s$ is necessary for
getting inflation, so the solution of (\ref{la_simple}) is approximately 
\be
	\lambda \cong {2 m_s^2\over 3 H_0}
\label{la2}
\ee  
Finally, matching coefficients at second order gives
\be
\s e^{\mu_1}+\LF\phi_2-\s\RF e^{\mu_2}=p\LF\phi_2-{p-1\over 2}\RF
\label{phi2}
\ee

In summary, by solving the scalar field equation of motion (\ref{KG}) to second order in the
expansion in powers of $u=e^{\lambda t}$, we have obtained two relations, 
 (\ref{la_simple}) and (\ref{phi2}).  The former determines the parameter $\lambda$ 
while the latter determines $\phi_2$.

\subsection{The Stress Energy Tensor and the Friedmann Equation}

To complete our approximate solution for the classical background, we must 
solve the Friedmann equation
\be
\label{Hubble}
H^2={1\over 3 M_p^2}\,\rho_{\phi}
\ee
to second order in $u$.
To find the energy density $\rho_{\phi}$, we turn to the 
stress energy tensor for the $p$-adic scalar field. 
A convenient expression for $T_{\mu\nu}$ was derived in \cite{stress_tensor} (see also 
\cite{Sen:stress})
\begin{eqnarray}
&&T_{\mu\nu} = {m_s^4\over 2g_p^2}g_{\mu\nu}\LT \phi e^{-{\Box\over m_p^2}}\phi
-{2\over p+1}\phi^{p+1}
+{1\over m_p^2}\int_0^1 d\tau\ \LF \Box e^{-{\tau\Box\over m_p^2}}\phi\RF\LF 
e^{-{(1-\tau)\Box\over m_p^2}}\phi\RF\right.\\
&&
\left.+{1\over m_p^2}\int_0^1 d\tau\ \LF \p_{\alpha} e^{-{\tau\Box\over m_p^2}}
\phi\RF\LF\p^{\alpha} e^{-{(1-\tau)\Box\over m_p^2}}\phi\RF\RT
-{m_s^4\over m_p^2g_p^2}\int_0^1 d\tau\ \LF \p_{\mu} 
e^{-{\tau\Box\over m_p^2}}\phi\RF\LF\p_{\nu} e^{-{(1-\tau)\Box\over m_p^2}}
\phi\RF 
 \nonumber 
\label{Tmunu}
\end{eqnarray}
One may verify that the $T_{\mu\nu}$ is symmetric by changing the dummy integration variable 
$\tau\rightarrow1-\tau$ in the last term.
For homogeneous $\phi(t)$ the above expression simplifies, and for $T_{00}$ we find
$$
\rho_{\phi}=-T_{00}={m_s^4\over 2g_p^2}\LT \phi e^{-{\Box\over m_p^2}}\phi-{2\over p+1}\phi^{p+1}+{1\over m_p^2}\int_0^1 d\tau\ \LF \Box e^{-{\tau\Box\over m_p^2}}\phi\RF\LF e^{-{(1-\tau)\Box\over m_p^2}}\phi\RF\right.
$$
\be
\left.+{1\over m_p^2}\int_0^1 d\tau\ \p_t\LF e^{-{\tau\Box\over m_p^2}}\phi\RF\p_t\LF e^{-{(1-\tau)\Box\over m_p^2}}\phi\RF\RT \label{rho}
\ee

One can evaluate  the above expression term by term, keeping up to 
$\cO(e^{2\la t})\sim u^2$. The final result reads
\begin{eqnarray}
T_{00} &=& {m_s^4\over 2g_p^2}\LT1-u(1+e^{\mu_1})-{2[1-(p+1)u]\over p+1}+u(e^{\mu_1}-1)\RT+\cO(u^2)
\nonumber \\
&=& {m_s^4(p-1)\over 2g_p^2(p+1)}+\cO(u^2)
\label{eden}
\end{eqnarray}
The $\cO(u)$ terms cancel out and matching the coefficients in the Friedmann equation gives us 
the simple results
\begin{equation}
H_0^2={m_s^4\over 6M_p^2}{p-1\over g_p^2(p+1)}
\label{H0}
\end{equation}
and
\begin{equation}
\label{H1}
  H_1 = 0
\end{equation}
for zeroth and first order respectively.

The  $\cO(u^2)$ contribution to $T_{00}$ is quite complicated (see appendix \ref{appa}) but 
once we use (\ref{H1}) it simplifies greatly.  Matching coefficient at order $\cO(u^2)$ 
in the Friedmann equation gives
\be
\label{H2}
H_2 = {\la m_s^4\over 4 g_p^2 m_p^2\,M_p^2}\,e^{\mu_1} 
    = \frac{1}{8 g_s^2}\frac{p^3 \ln p}{p-1}\left(\frac{m_s}{M_p}\right)^2\lambda
\ee
Because of our sign convention for $H_r$, the fact that $H_2>0$ means that the expansion
is slowing as $\phi$ rolls from the unstable maximum, as one would expect in a conventional
inflationary model.  

Using (\ref{la2}), (\ref{H0}) and (\ref{H2}) to compute  $H_2 / H_0 \sim p \ln p$ it is clear
that the perturbative expansion in $u$ breaks down once $u \sim (p\ln p)^{-1/2}$ (recall that
$H \cong H_0 - H_2 u^2$).  Thus one expects that once $u \sim p^{-1/2}$ then inflation ends.
We verify this claim using an alternative formalism in the next subsection.

To summarize, we have determined the five parameters  $\la, \phi_2, H_0,  H_1$ and $H_2$ 
which appear in the solutions for $\phi(t)$ and $H(t)$ up to 
 $\cO(u^2)$ through  the equations (\ref{la_simple}-\ref{phi2}), (\ref{H0}-\ref{H2}). 
As a check of our result, we can take $H_0\ll m_s$ and compare it to the 
the Minkowski background solution that was found 
in \cite{zwiebach}.  For $H_0\ll m_s$ we 
see from (\ref{la_simple}) that 
\be
 \la\cong \sqrt{2} m_s\LF1-{3H_0\over 2\sqrt{2}m_s}\RF
\ee
the first term corresponding to the known Minkowski result.
We can also compute the coefficient $\phi_2$ in this limit from (\ref{phi2}),
\be
\phi_2\cong -{1\over 2(p^2+p+1)}
\ee
This too coincides with the coefficient that was determined in \cite{zwiebach} for $p=2$. 


\subsection{The Friction-Dominated Approximation}



In the previous subsections we have constructed an approximate solution for the $p$-adic
scalar rolling down its potential by performing an expansion of $\phi$, $H$ in a power series
in $u$.  Furthermore, we have shown that once $u \sim p^{-1/2}$ then this solution breaks
down (because the $\mathcal{O}(u^2)$ term in $H(t)$ become larger than the zeroth order term, 
$H_0$).  At this point inflation has ended.  Because the equations of motion are complicated, we 
now verify this behaviour using an alternative formalism which does not rely on small $u$.

The method is the same as the slow-roll approximation in ordinary inflation,
which assumes 
that $\dot\phi\ll H\phi$.  To justify it within the $p$-adic theory,  we will 
provisionally assume that 
\[
  \lambda^2 \ll 3 H_0 \lambda
\]
so that the evolution is friction-dominated in the usual sense.  
The consistency of this approximation will be established when we match the theory
to the observables from the inflationary power spectrum later, 
in eqs.\ (\ref{msh}), (\ref{lambda}). It follows that it is
a good approximation to take
\[
  -\Box \cong 3 H_0 \partial_t
\]
Then the $p$-adic scalar field equation becomes
\begin{equation}
\label{friction_KG}
  e^{\alpha \partial_t}\phi = \phi^p
\end{equation}
where we have defined
\begin{equation}
\label{alpha}
  \alpha \equiv \frac{3 H_0}{m_p^2}
\end{equation}
Our procedure is to treat $H_0$ as exactly constant, solve for $\phi(t)$ and compute the 
energy density $\rho_\phi$.  If $\rho_\phi$ is approximately constant then this series of 
approximations
is self-consistent and the solution is reliable.  Once $\rho_\phi$ begins to deviate significantly
from a constant value then the solution breaks down and we conclude that inflation has ended.

We now proceed to solve (\ref{friction_KG}) for constant $H_0$.  To this end we expand 
$\phi(t)$ in Fourier modes as
\begin{equation}
\label{fourier}
  \phi(t) = \frac{1}{\sqrt{2\pi}} \int_{-\infty}^{+\infty} dk e^{-ikt} \phi_k
\end{equation}
so that
\begin{eqnarray}
  e^{\alpha \partial_t} \phi(t) &=& 
  \frac{e^{\alpha \partial_t}}{\sqrt{2\pi}} \int_{-\infty}^{+\infty} dk e^{-ikt} \phi_k \\
  &=&
  \frac{1}{\sqrt{2\pi}} \int_{-\infty}^{+\infty} dk e^{-ik(t+\alpha)} \phi_k \\
  &=& \phi(t+\alpha)
\end{eqnarray}
The $p$-adic scalar equation of motion (\ref{friction_KG}) then takes the simple form
\begin{equation}
\label{shift_eqn}
  \phi(t+\alpha) = \phi(t)^p 
\end{equation}
It is quite remarkable that this equation admits dynamical solutions.  It is straightforward
to check that (\ref{shift_eqn}) has the solution
\begin{equation}
\label{shift_soln}
  \phi(t) = e^{-u(t)}
\end{equation}
where $u(t) = e^{\lambda t}$ and $\lambda = \alpha^{-1} \ln p = 2 m_s^2 / (3 H_0)$, as before.
One can easily check by acting on (\ref{shift_soln}) with the full operator 
$-\Box = \partial_t^2 + 3H_0\partial_t$ that indeed the friction term dominates as long as 
$u \lsim 1$.

Performing a Taylor expansion of (\ref{shift_soln}) about $u=0$ gives
\[
  \phi \cong 1 - u + \frac{1}{2} u^2 + \cdots
\]
which reproduces our solution in the small-$u$ expansion in the limit that 
$m_s \ll H_0$\footnote{We will see later that this is the same as taking the spectral
index equal to unity $n_s \rightarrow 1$.}
(see equations \ref{2nd_order_expasions} and \ref{phi2_soln}).  We also have
\[
  e^{-\Box / m_p^2}\phi = e^{-up} \cong 1 - p u + \frac{1}{2}p^2 u^2
\]
which again reproduces our previous results (which can be verified by inserting the solutions
for $\mu_1$, $\mu_2$, $\sigma$ and $\phi_2$ into equation \ref{eboxphi}).

We now proceed to construct the energy density for $\phi(t)$ in this approximation.  It is 
straightforward to show that
\begin{eqnarray}
  e^{-\tau\Box/m_p^2}\phi &=& \exp\left(-u p^{\tau}\right) \nonumber \\
  e^{-(1-\tau)\Box/m_p^2}\phi &=& \exp\left(-u p^{1-\tau}\right) \label{tau_terms}
\end{eqnarray}
We write $\rho_\phi$ (\ref{rho}) as
\begin{equation}
\label{break_into_terms}
  \rho_\phi \equiv \frac{m_s^4}{2 g_p^2}\left[ T_1 + T_2 + T_3 + T_4 \right]
\end{equation}
where
\begin{eqnarray}
  T_1 &\equiv& \phi e^{-\Box/m_p^2}\phi \label{T1} \\
  T_2 &\equiv& -\frac{2}{p+1}\phi^{p+1} \label{T2} \\
  T_3 &\equiv& \frac{1}{m_p^2}\int_0^1d\tau 
          \left(\Box e^{-\frac{\tau\Box}{m_p^2}}\phi \right)
          \left(e^{-\frac{(1-\tau)\Box}{m_p^2}}\phi\right) \label{T3} \\
  T_4 &\equiv& \frac{1}{m_p^2}\int_0^1d\tau \partial_t\left(e^{-\frac{\tau\Box}{m_p^2}}\phi\right)
               \partial_t\left(e^{-\frac{(1-\tau)\Box}{m_p^2}}\phi\right) \label{T4}
\end{eqnarray}
Using the scalar field equation (\ref{KG}) and (\ref{shift_soln}) the first two terms, (\ref{T1})
and (\ref{T2}), are trivial
\begin{equation}
\label{T1+T2}
  T_1 + T_2 = \frac{p-1}{p+1}e^{-up-u}
\end{equation}
Since this term is proportional to $\phi^{p+1}$ we identify it with the potential energy
of the $p$-adic scalar.

The next simplest term to evaluate is $T_4$ (\ref{T4}) which gives
\[
  T_4 = \frac{\lambda^2 u^2}{2m_s^2} p\ln p \int_0^1d\tau
  \exp\left[-u\left(p^{\tau} + p^{1-\tau}\right)\right]
\]
It is useful to change the variable of integration to $\omega = u p^{\tau}$ and cast this
result in the form
\begin{equation}
\label{T4_soln}
  T_4 = \frac{\lambda^2}{2m_s^2} ( p u^2 ) \int_{u}^{up}\frac{d\omega}{\omega}
  \exp\left[-\omega - \frac{p u^2}{\omega}\right]
\end{equation}
The $d\omega$ integral can be performed exactly (though not in closed form) in terms of
special functions.  Since $T_4$ is subdominant to a contribution coming from $T_3$
we will not investigate the behaviour of (\ref{T4_soln}) any further.

We now study $T_3$ (\ref{T3}).  This can be written in the form
\begin{eqnarray}
  T_3 &=& -\frac{\lambda^2}{2m_s^2}\int_{u}^{up}d\omega \omega
  \exp\left[-\omega - \frac{p u^2}{\omega}\right] \nonumber \\
  &+& \frac{\lambda^2}{2m_s^2}\left(1 + \frac{3H_0}{\lambda}\right)
       \int_{u}^{up}d\omega 
  \exp\left[-\omega - \frac{p u^2}{\omega}\right] \label{T3_soln}
\end{eqnarray}
The dominant contribution to $T_3$ is the one proportional to $H_0$ since the evolution is 
friction-dominated.  This term is also larger than $T_4$.  The leading contribution to 
$T_3 + T_4$ is then
\begin{equation}
\label{T3+T4}
  T_3 + T_4 \cong \int_{u}^{up}d\omega 
  \exp\left[-\omega - \frac{p u^2}{\omega}\right]
\end{equation}
where we have used the fact that $3 H_0 \lambda / (2 m_s^2) \cong 1$ which follows from 
(\ref{la_simple}) when $\lambda^2 \ll 3 H_0 \lambda$.  Since $T_3$ and $T_4$ contain time
derivatives acting on $\phi$ it is natural to identify (\ref{T3+T4}) with the kinetic energy
of the $p$-adic scalar.

Let us study the behaviour of the kinetic energy, equation (\ref{T3+T4}), as a function of $u$.
The integral in (\ref{T3+T4}) can be performed in terms of the incomplete cylindrical functions
of the Sonine-Schlaefli form \cite{bessel} (see appendix \ref{appb} for a review).
\[
  \int_{u}^{up}d\omega 
  \exp\left[-\omega - \frac{p u^2}{\omega}\right] 
  = 2\pi u p^{1/2} S_{-1}(-u,-up; 2iup^{1/2})
\]
We now study the behaviour of this integral in various limits.  We assume that 
$p \gg 1$
throughout since the previous method of expanding in $u$ is valid in the case where
$p\sim 1$.  At very early times, $u < 1/p$, this integral goes to zero as
\[
  \int_{u}^{up}d\omega 
  \exp\left[-\omega - \frac{p u^2}{\omega}\right] \cong u p 
  \hspace{5mm}\mathrm{for}\hspace{5mm}u < 1/p
\]
For intermediate times, $1/p \ll u \lsim 1/p^{1/2}$, it is a good approximation to treat the upper
and lower limits of integration as $0$ and $+\infty$ respectively.  In this approximation we
can write the incomplete cylinder function in terms of a Hankel function of order 
$\nu = -1$ (see appendix \ref{appb} for details).  The small argument asymptotics of 
$H_{-1}^{(1)}$ gives
\[
  \int_{u}^{up}d\omega 
  \exp\left[-\omega - \frac{p u^2}{\omega}\right] \cong 1 + 2 u^2 p \ln(2up^{1/2}) 
  \hspace{5mm}\mathrm{for}\hspace{5mm}1/p \ll u < 1/p^{1/2}
\]
Finally we consider late times, $u > 1/p^{1/2}$.  It is still reasonable to extend the integral
as $\int_u^{up}d\omega \cong \int_0^\infty d\omega$ and hence the integral can still be written
in terms of $H_{-1}^{(1)}$.  This time the large-argument asymptotics of the Hankel function 
are appropriate and one has
\[
  \int_{u}^{up}d\omega 
  \exp\left[-\omega - \frac{p u^2}{\omega}\right] \cong \sqrt{\pi u p^{1/2}} e^{-2up^{1/2}}
  \hspace{5mm}\mathrm{for}\hspace{5mm}u > 1/p^{1/2}
\]
We have verified these asymptotic expressions numerically.

We can now write the dominant contribution to $\rho_\phi$ in the 
friction-dominated approximation,
\begin{equation}
\label{friction_rho}
  \rho_\phi = \frac{m_s^4}{2g_s^2}\frac{p^2}{p-1}
  \left(\frac{p-1}{p+1}e^{-up - u} + \int_{u}^{up}d\omega 
  \exp\left[-\omega - \frac{p u^2}{\omega}\right]  \right)
\end{equation}
The first term, proportional to $e^{-up}$, represents the potential energy and the
second term represents the kinetic energy.  Using our previous analysis of the
kinetic energy the  behaviour of $\rho_\phi$ as a function of $u$ (assuming $p\gg1$)
is clear.  At early times  $u < 1/p$ the potential energy dominates and we have
$\rho_\phi \cong m_s^4/(2g_p^2)$.   At intermediate times $1/p \ll u \lsim
1/p^{1/2}$ the potential energy goes to zero and the  kinetic energy dominates and
we have $\rho \cong m_s^4/(2g_p^2)$.  At late times $u > 1/p^{1/2}$  and $\rho_\phi$
damps to zero as $e^{-2up^{1/2}}$.  We have verified this behaviour numerically.
Figure \ref{rhofig1} shows the behaviour of $\rho_\phi$ as a function of $t$, 
verifying that $\rho_\phi$ is approximately constant for $u < p^{-1/2}$.   The  time
evolution of both the potential energy and the kinetic energy contributions to 
$\rho_\phi$ are shown, demonstrating that the latter dominates in the interval
$p^{-1} < u < p^{-1/2}$.   In this figure we have taken $p = 10^5$ for illustrative
purposes.

\EPSFIGURE[ht]{padic_rho3.eps,width=4in}{
Plot of the energy density of the $p$-adic scalar as a function of $t$,
for $p = 10^5$.
This figure shows that $\rho_\phi$ is
approximately constant for $u < p^{-1/2}$\label{rhofig1}.  The individual
contributions from the potential
and kinetic energy are also shown.}

We conclude that for $u \lsim 1/p^{1/2}$ the scalar field
energy density $\rho_\phi$ is approximately constant and inflation proceeds.  At $u = 1/p^{1/2}$
the energy density of the $p$-adic scalar begins to decrease quickly and the analysis of this 
subsection is no longer applicable.  We conclude that inflation ends, roughly, when $u=1/p^{1/2}$.

It is quite interesting that during the intermediate phase $p^{-1} < u < p^{-1/2}$ it is in fact
the \emph{kinetic} energy which is driving inflation, rather than the potential energy.  This
is quite different from what occurs in a local field theory.

\section{Fluctuations and Inflationary Predictions}
\label{fluct}

In this section we consider the spectrum of cosmological fluctuations produced during
$p$-adic inflation.  The full cosmological perturbation theory for the $p$-adic string model (\ref{action}), which 
should include metric perturbations and also take into account the departure of the background 
expansion from pure de Sitter, is complicated and beyond the scope of the present paper. We
leave a detailed study of these matters to future investigation \cite{inprog}.  To  simplify
the analysis we will  neglect scalar metric perturbations as well as deviations of the
background metric from pure de Sitter space during horizon crossing.  In standard
cosmological perturbation theory these approximations reproduce the more exact results to
reasonable accuracy and we therefore assume that the situation is similar for the action
(\ref{action}).

\subsection{$p$-adic Tachyon Fluctuations}

We are approximating the background dynamics as de Sitter which amounts to working in the
limit $u \rightarrow 0$ so that
\begin{eqnarray}
  H^2 &\equiv& H_0^2 = \frac{m_s^4}{6M_p^2}\frac{p-1}{g_p^2(p+1)} \\
  \phi &\equiv& \phi_{0} \equiv 1
\end{eqnarray}
We expand the $p$-adic tachyon field in perturbation theory as
\begin{eqnarray*}
  \phi(t,\vec{x}) &=& \phi^{(0)}(t) + \delta \phi(t,\vec{x}) \\
                  &=& 1 + \delta \phi(t,\vec{x})
\end{eqnarray*}
The perturbed Klein-Gordon equation (\ref{KG}) takes the form
\begin{equation}
\label{pert_KG}
  e^{-\Box/ m_p^2}\, \delta \phi = p \,\delta \phi
\end{equation}
(Inhomogeneous solutions in $p$-adic string theory have also been considered in
\cite{inhomo}.)
One can construct solutions by taking $\delta \phi$ to be an eigenfunction of the $\Box$
operator.  If we choose $\delta \phi$ to satisfy
\begin{equation}
\label{eigenvalue}
  -\Box \delta\phi = + B\, \delta \phi
\end{equation}
then this is also a solution to (\ref{pert_KG}) if
\begin{equation}
\label{B}
  B = m_p^2 \ln p = 2 m_s^2
\end{equation}
where in the second equality we have used (\ref{gpmp}).

The solutions of (\ref{eigenvalue}) are well known.  However, in order to make contact 
with the usual treatment of cosmological perturbations we need to define a field
in terms of which the action appears canonical.  This presents a serious difficulty because,
in general, there is no local field redefinition which will bring the kinetic term 
$\phi \left(1 - e^{-\Box/ m_p^2}\right) \phi$ into the canonical form $\phi \Box \phi$.
(One might imagine simply truncating the expansion in powers of $\Box$ as we have described
in section \ref{sect3}, however, the higher order terms are not negligible in general.)
Fortunately, for fields which are on-shell (that is, when (\ref{eigenvalue}) is solved)
the field obeys
\begin{eqnarray*}
  \left(1 - e^{-\Box/ m_p^2}\right) \delta \phi &=& \left(1-e^{B/m_p^2}\right)\delta \phi \\
  &=&  \left(1-e^{B/m_p^2}\right)\frac{1}{(-B)}\,(-B)\delta \phi \\
  &=& \left(1-e^{B/m_p^2}\right)\frac{1}{(-B)}\,\Box\delta \phi \\
  &=& \frac{p-1}{2m_s^2}\,\Box\delta \phi 
\end{eqnarray*}
Thus, for on-shell fields the kinetic term in the Lagragian can be written as
\begin{eqnarray}
  \mathcal{L}_{\mathrm{on-shell}} &=& \frac{m_s^4}{g_p^2}\frac{1}{2}\phi (1-e^{-\Box/m_p^2})\phi 
  + \cdots \nonumber \\
  &=& \frac{m_s^4}{g_p^2}\frac{p-1}{2m_s^2}\frac{1}{2}\phi\Box\phi + \cdots \nonumber \\
  &=& \frac{1}{2}\varphi \Box \varphi + \cdots \label{can}  
\end{eqnarray}
In (\ref{can}) we have defined the ``canonical'' field
\begin{equation}
\label{varphi}
  \varphi \equiv A \phi
\end{equation}
where
\begin{equation}
\label{A}
  A \equiv \frac{m_s p}{\sqrt{2}\, g_s}
\end{equation}
The field $\varphi$ has a canonical kinetic term in the action, at least while 
(\ref{eigenvalue}) is satisfied.  Notice that $\varphi$ is distinct from the field
$\chi$ (see (\ref{canonical})) which we introduced in section \ref{sect3}.
The field $\chi$ corresponds to the canonical field which one would naively define when
neglecting terms $\mathcal{O}(\Box^2)$ and higher in the action (as is typical in
studies of tachyonic inflation) while $\varphi$ is the appropriate definition of 
the canonically normalized field when taking into account the infinite series of 
higher derivative corrections.

Now, let us return to the task of solving (\ref{eigenvalue}), bearing in mind that
$\delta\varphi = A \delta \phi$ is the appropriate canonically normalized field.
We write the quantum mechanical solution in term of annihilation/creation operators as
\[
  \delta \varphi(t,\vec{x}) = \int \frac{d^{\,3}k}{(2\pi)^{3/2}}\left[a_k \varphi_k(t) e^{ikx} + 
                            \mathrm{h.c.} \right]
\]
and the mode functions $\varphi_k(t)$ are given by
\begin{equation}
\label{mode}
  \varphi_k(t) = \frac{1}{2}\sqrt{\frac{\pi}{a^3 H_0}}\,e^{\frac{i\pi}{2}\left(\nu + 1/2\right)}
            \, H_\nu^{(1)}\!\left(\frac{k}{a H_0}\right)
\end{equation}
where the order of the Hankel functions is
\begin{equation}
\label{nu}
  \nu = \sqrt{\frac{9}{4} + \frac{B}{H_0^2}} = \sqrt{\frac{9}{4} + \frac{2 m_s^2}{H_0^2}}
\end{equation}
and of course $a = e^{H_0 t}$.  In the second equality in (\ref{nu}) we have used
(\ref{B}) and (\ref{gpmp}).  In writing (\ref{mode}) we have used the usual Bunch-Davies
vacuum normalization so that on small scales, $k \gg a H_0$, one has
\[
  |\varphi_k| \cong \frac{a^{-1}}{\sqrt{2k}}
\]
which reproduces the standard Minkowski space fluctuations.  This is the usual procedure in
cosmological perturbation theory.  However, we note that the quantization of the theory
(\ref{action}) is not transparent and it  might turn out that the usual
prescription is incorrect in the present context.  We defer this and other subtleties to future
investigation.\footnote{Our prescription for choosing the vacuum has the property that
in the local limit $p\rightarrow 1$ the cosmological fluctuations are identical
to the well-known solutions in local field theory.}  On large scales, $k \ll a H_0$, 
the solutions (\ref{mode}) behave as
\[
  |\varphi_k| \cong \frac{H}{\sqrt{2k^3}}\left(\frac{k}{a H_0}\right)^{3/2-\nu}
\]
which gives a large-scale power spectrum for the fluctuations
\[
  P_{\delta\varphi} = \left(\frac{H_0}{2\pi}\right)^2\left(\frac{k}{a H_0}\right)^{n_s-1}
\]
with spectral index
\[
  n_s - 1 = 3 - 2\nu
\]
From (\ref{nu}) it is clear that to get an almost scale-invariant spectrum we require
$m_s \ll H_0$.  In this limit we have
\begin{equation}
\label{n}
  n_s - 1 \cong -\frac{4}{3} \left(\frac{m_s}{H_0}\right)^2
\end{equation}
which gives a red tilt to the spectrum, in agreement with the latest WMAP data
\cite{wmap}.  For
$n_s \cong 0.95$ one has $m_s \cong 0.2 H_0$.  Comparing (\ref{mode}) to the corresponding
solution in a local field theory we see that the $p$-adic tachyon field fluctuations evolve
as though the mass-squared of the field was $-2 m_s^2$ which may be quite different from
the mass scale which one would infer by truncating the infinite series of derivatives: 
$\partial ^2 V / \partial \chi^2 (\chi = \chi_0)$ (see eq.\ (\ref{mass})).  
The fact that $H_0 > m_s$ is an unusual feature; we will comment on it below.

It is worth pointing out that we have constructed solutions of a partial differential
equation with infinitely many derivatives, for which we are free to specify two initial
data; to obtain eq.\ \ref{mode} we fixed these using the Bunch-Davies prescription.
Precisely the same result was obtained when inhomogeneous solutions were studied in 
\cite{inhomo}.  Partial differential equations with infinitely many derivatives constitute a
new class of equations in mathematical physics about which little is presently known. In
particular, it is not clear how to pose the intial value problem for such equations. We
conjecture that the most general solutions of (\ref{KG}) are specified by two initial data
(for example $\phi(0,\vec{x})$ and $\dot{\phi}(0,\vec{x})$), just like equations containing
only one power of $\Box$.  See \cite{math} for mathematical work on constructing solutions
of  equations with infinitely many derivatives. 

\subsection{Determining Parameters} 

We now want to fix the parameters of the model by comparing to the observed features
of the CMB perturbation spectrum. There are three dimensionless
parameters, $g_s,\ p$ and the ratio $m_s/M_p$. The important question is whether there is
a sensible parameter range which can account for CMB observations, \ie, the spectral tilt and
the amplitude of fluctuations.  Using (\ref{H0}) in (\ref{n}), we can relate the tilt
to the model parameters via
\begin{equation}
\label{string_scale} 
	|n_s-1| =
	\frac{8(p+1)}{p^2}\left(\frac{M_p}{m_s}\right)^2g_s^2 \quad\im\quad
	\left(\frac{m_s}{M_p}\right)^2=\frac{8(p+1)}{p^2}\frac{g_s^2}{|n_s-1|} 
\end{equation} 
Thus one
can have  a small tilt while ensuring that the string scale is smaller than the
Planck scale, provided that $g_s^2/p \ll 1$.  Henceforth we will use (\ref{string_scale})
to determine $m_s/M_p$  in terms of $p$, $g_s$, and $|n_s-1|\cong 0.05$.
All the dimensionless parameters in our solution, 
$m_s/H_0$, $\lambda/m_s$, $\phi_2$ and $H_2/m_s$,  are likewise functions 
of $n_s-1, p$ and $g_s$.   From (\ref{H0}) and (\ref{string_scale}) we see that for $p\gg 1$,
\be
	{m_s\over H_0} \ \cong\ \sqrt{6}\, g_p {M_p\over m_s} \ \cong\ g_s \sqrt{{6\over p}}\, 
 {M_p\over m_s} \ \cong\ \frac12\sqrt{3|n_s-1|}
\label{msh}
\ee
It may seem strange to have  $H$ exceeding $m_s$ since
that means the energy density exceeds the fundamental scale, but this is an inevitable 
property of the $p$-adic tachyon at its maximum, as shown in eq.\ (\ref{eden}).  This
is similar to other attempts to get tachyonic or brane-antibrane inflation from string theory, since the
false vacuum energy is just the brane tension which goes like $m_s^4/g_s$.  

Next we determine $\lambda/m_s$, where $\lambda$ is the mass scale appearing in the
power series in $e^{\lambda t}$ which provides the ansatz for the background solutions.
Consider eq.\  (\ref{la}) for $\lambda$ in the $H_0\gg m_s$ limit.  
The positive root for $\lambda$ gives
\begin{equation}
\label{lambda}
  \frac{\lambda}{m_s} \cong \sqrt{\frac{|n_s-1|}{3}}
\end{equation} 
From (\ref{msh}) and (\ref{lambda}) we find that $3 H_0 \lambda \gg \lambda^2$ which means that 
the evolution is friction-dominated in the usual sense.  
As for $\phi_2$, from  
eq.\ (\ref{phi2}) it follows that
\begin{equation}
\label{phi2_soln}
  \phi_2 \cong -\frac{1}{2}\frac{p-1}{p^{1+\beta}-1}
\end{equation}
with 
\[
\beta \cong |n_s-1| / 3 + \mathcal{O}(|n_s-1|^2) 
\]
so that $\beta \ll 1$.  Notice that for $p \gg 1$, we have $\phi_2 \cong -0.5 p^{-\beta}$.
Finally we have $H_2$ which from eq.\ (\ref{H2}) is given by
\begin{equation}
\label{H2_soln}
  \frac{H_2}{m_s} = \frac{p}{\sqrt{3}}\frac{p-1}{p+1}\frac{\ln p}{|n_s-1|^{1/2}}
\end{equation}
To go further, we must impose the COBE normalization on the amplitude of the density
perturbations, to show that
it is possible to satisfy all the experimental constraints while keeping 
$m_s/M_p<1$.  The latter requirement is usually needed for the validity of any 4D
effective description of string theory.  
In a compactification of the $d$ extra dimensions whose volume is of
order $V_d\sim m_s^{-d}$, we would have $M_p\sim m_s$ whereas more generally
$M_p^2 = m_s^{2+d} V_d$.  The 4D effective theory would normally need to be supplemented by
higher dimension operators if $V_d$ was small compared to $m_s^{-d}$.

\subsection{Curvature Perturbation and COBE normalization}

In order to fix the amplitude of the density perturbations we consider the curvature
perturbation $\zeta$.  We assume that
\[
  \zeta \sim -\frac{H}{\dot{\varphi}} \delta \varphi
\]
as in conventional inflation models.  
To evaluate the prefactor $H/\dot{\varphi}$ we must work beyond
zeroth order in the small $u$ expansion.  We take $\phi = 1 - u$ to evaluate the prefactor,
even though the perturbation $\delta\chi$ is computed in the limit that $\phi = 1$.  This
should reproduce the full answer up to $\mathcal{O}(u)$ corrections.
The prefactor is
\begin{eqnarray*}
  -\frac{H}{\dot{\varphi}} &\cong& \frac{H_0}{A \lambda u} \\
   &\cong& \frac{2^{3/2} g_s}{p}\,\frac{1}{|n_s-1|}\,\frac{1}{u}\,m_s^{-1}
\end{eqnarray*}
We should evaluate $u$ at the time of horizon crossing, $t_\star$, defined to be
approximately 60 e-foldings before the end of inflation 
$t_{\mathrm{end}}$, assuming that the energy scale of inflation is high (near the GUT scale).
We must therefore estimate $t_{\mathrm{end}}$.  We have shown in the last subsection that
inflation ends when $u \sim 1/p^{1/2}$.
From eqs.\ (\ref{msh}-\ref{lambda}) we see that $H_0/\lambda = 2/|n_s-1|$;
therefore we can write the scale factor
$a(t) \cong e^{H_0 t}$ in the form
\be
  a(t) \cong u(t)^{2/|n_s-1|}
\ee
so that $a_\star = e^{-60} a_{\mathrm{end}}$ corresponds to
\be
  u_\star = e^{-30|n_s-1|} u_{\mathrm{end}} \equiv e^{-30|n_s-1|}\frac{1}{p^{1/2}}
\label{ustar}
\ee

The power spectrum of the curvature perturbation is given by
\begin{equation}
\label{P_zeta}
  P_\zeta = \left|\frac{H}{\dot{\varphi}}\right|^2 P_{\delta\varphi} \equiv 
  A_\zeta^2 \left(\frac{k}{a H_0}\right)^{n_s-1}
\end{equation}
where the amplitude of fluctuations $A_\zeta$ can now be read off as
\begin{equation}
\label{scalar_amp}
  A_\zeta^2 = \frac{8}{3\pi^2}\frac{g_s^2}{p}\frac{e^{60|n_s-1|}}{|n_s-1|^3}
\end{equation}
As an example, taking $n_s \cong 0.95$ one can fix the amplitude of the density perturbations 
$A_\zeta^2 \cong 10^{-10}$ by choosing
\begin{equation}
\label{cobe}
  \frac{g_s}{\sqrt{p}} \cong 0.48 \times 10^{-7}
\end{equation}

To get a more general idea of how the inflationary observables constrain the
parameters of the model, we will allow $n_s$ to vary away from the value $0.95$, which
is a fit to the WMAP data under the assumption that the tensor contribution to the
spectrum is negligible.  Setting $A_\zeta^2 = 10^{-10}$ and using (\ref{scalar_amp})
gives and expression for $g_s$ in terms of $p$ and $|n_s-1|$
\be
g_s = \sqrt{ \frac{3\pi^2}{8}}\, \sqrt{p}\, e^{-30|n_s-1|} |n_s-1|^{3/2} \times 10^{-5}
\label{gs}
\ee
Combining (\ref{gs}) with (\ref{string_scale}), we also obtain
\be
{m_s\over M_p} = \sqrt{ 3\pi^2 }\,\sqrt{\frac{p+ 1 }{p}}\, e^{-30|n_s-1|} |n_s-1| \times 10^{-5}
\label{ms}
\ee
We graph the dependence of $g_s$ and $m_s/M_p$ on $p$ for several  values of
the spectral index in figures \ref{gsfig} and \ref{msfig}.  We see that the string scale
is bounded from above as $m_s / M_p \lsim 0.94 \times 10^{-6}$ and that for typical values
of $p,n_s$ it is close to $m_s / M_p \cong 0.61 \times 10^{-6}$.  We also see from (\ref{gs})
that $g_s$ is unconstrained and that $g_s$, $p$ are not independent parameters.  If we wish
to take $g_s \sim \mathcal{O}(1)$ then we must choose extremely large values $p \sim 10^{14}$.
If we restrict ourselves to the perturbative regime $g_s < 1$ then this places an upper bound
on $p$:
\[
  p < 4.3 \times 10^{14}
\]

\DOUBLEFIGURE[ht]{gs2.eps,width=\hsize}{ms2.eps,width=\hsize}
{\label{gsfig} Log of the string coupling $g_s$ as a function of $\log_{10} p$
for several values of $n_s$.} {\label{msfig} Log of the ratio of the string mass 
scale to the Planck mass as a function of $\log_{10} p$
for several values of $n_s$.}

\subsection{Comments on Slow Roll and the Relation to Local Field Theory}
\label{SRsubsect}

It is quite remarkable that our predictions for inflationary observables and also the
solutions for $\chi$, $H$ are essentially identical to the results from \emph{local} field 
theory with potential $V(\psi) = g_s(\psi^2 - v^2)^2/4$ where $g_s v^2 \equiv 2 m_s^2$ 
(see appendix \ref{appc} for a detailed comparison).
Indeed, the dynamics of the $p$-adic tachyon are fixed by the mass scale in 
the kinetic term, $m_s$, rather than by the naive mass scale 
$\sqrt{-\partial^2V/\partial\chi^2}$ (see section \ref{sect3}), which may be much larger than 
$m_s$.  

It is an interesting feature of this theory that the canonical $p$-adic tachyon, $\phi$,
can roll slowly despite the fact that, working in a derivative truncatation (as in section 
\ref{sect3}), one would conclude that the tachyon has an extremely steep potential.  
To see this we first define the Hubble slow roll parameters $\epsilon_\sH$, $\eta_\sH$ by
\begin{eqnarray}
  \epsilon_\sH &\equiv& \frac{1}{2M_p^2} \frac{\dot{\varphi}^2}{H^2}\label{epsilon} \\
  \epsilon_\sH - \eta_\sH &\equiv& \frac{\ddot{\varphi}}{H \dot{\varphi}} \label{eta}
\end{eqnarray}
These are the appropriate parameters to describe the rate of time variation of the inflaton 
as compared to the Hubble scale.  Using the solution $\phi \cong 1-u$ 
(recall that $\varphi = A\phi$, $A = m_s p / (\sqrt{2}\,g_s)$) we find that
\begin{eqnarray}
  \epsilon_\sH &\cong& \frac{1}{2}\frac{p+1}{p} e^{-60|n-1|}|n_s-1| 
\label{eH}                         \\
   \eta_\sH &\cong& - \frac{|n_s-1|}{2} 
\end{eqnarray}
We see that the Hubble slow-roll parameters, as defined above, are small.  This means
that the $p$-adic tachyon field rolls slowly in the conventional sense.  One reaches the same
conclusion if one defines the potential slow roll parameters 
\emph{using the correct canonical field}, which is $\varphi$ (\ref{varphi}):
\begin{eqnarray}
  \frac{M_p^2}{2}\left.   
\left(\frac{1}{V}\frac{\partial V}{\partial \varphi}\right)^2\right|_{\varphi = A} &=& 0 \\
  M_p^2 \left.\frac{1}{V}
  \frac{\partial^2 V}{\partial\varphi^2}\right|_{\varphi = A} &=& -\frac{1}{2}|n_s-1|
\end{eqnarray}

On the other hand, consider the potential slow roll parameter which one would naively define
using the the derivative truncated action (\ref{local}):
\begin{eqnarray}
   \frac{M_p^2}{2}\left. \left(\frac{1}{V}\frac{\partial V}{\partial \chi}\right)^2\right|_{\chi = \chi_0} &=& 0 \label{epsilon_V} \\
  M_p^2 \left. \frac{1}{V}\frac{\partial^2 V}{\partial \chi^2} \right|_{\chi = \chi_0}
  &=& - \left(\frac{p-1}{\ln p}\right)\, \frac{1}{2}|n_s-1|\label{eta_V}
\end{eqnarray}
where in (\ref{eta_V}) we have used equations (\ref{V_curv}) and (\ref{string_scale}).
We see that (\ref{eta_V}) can be enormous, though the tachyon field rolls slowly.
Taking the largest allowed value of $p$, $p \sim 10^{14}$, and $n_s \cong 0.95$ we
have $M_p^2 V^{-1}|\partial^2 V/\partial \chi^2| \sim 10^{11}$!  

Since large values of $p$ are required  if one wants to obtain $g_s \sim 1$, it follows that it is
somewhat natural for $p$-adic inflation to operate in the regime where the higher derivative
corrections play an important role in the dyanimcs.
However, in the regime where  $p\sim 1$ (corresponding to very small coupling $g_s$)
this novel features is not present. For example, with $p = 3$, $n_s=0.95$ one has
$g_s \sim 10^{-7}$ and $M_p^2 V^{-1}|\partial^2 V/\partial \chi^2| \sim 0.05$
and so the slow-roll dynamics are not surprising.

\subsection{Tensor Modes}

Since the $p$-adic stress tensor (\ref{Tmunu}) does not contribute any anisotropic stresses
up to first order in perturbation theory it follows that the first order tensor perturbations 
of the metric do not couple to the first order tachyon perturbation 
(see, for example, \cite{review}).  In fact, the action for the tensor perturbations
is given by
\[
  S_{\mathrm{grav}} = \frac{M_p^2}{2}\int dt\, d^{\,3}x\, a(t)^3\, \frac{1}{2} 
                      \partial_\mu h_{ij}\, \partial^\mu h^{ij}
\]
The standard procedure gives a power spectrum for the gravity waves
\[
  P_{T} = A_T^2\left(\frac{k}{a H_0}\right)^{n_T}
\]
with amplitude
\begin{equation}
\label{tensor_amp}
  A_T^2 = \frac{8}{M_p^2}\left(\frac{H_0}{2\pi}\right)^2
\end{equation}
We would like to compare this to the power in scalar modes (\ref{scalar_amp}).  
Defining the tensor-to-scalar ratio in the usual way
\[
  r \equiv \frac{\frac{1}{100}A_T^2}{\frac{4}{25}A_\zeta^2}
\]
we find that
\begin{equation}
\label{r}
  r = \frac{1}{2 M_p^2}\frac{\dot{\varphi}^2}{H_0^2} = \epsilon_\sH
\end{equation}
which reproduces the usual result from local field theory.  Using (\ref{eH}), we can evaluate 
$r$ as a function of $p$ and $n_s-1$
\begin{equation}
\label{r2}
  r = \frac{1}{2}\frac{p+1}{p} e^{-60|n_s-1|} |n_s-1|
\end{equation}
It is easy to see that $r$ is maximal when $p = 2$, $n_s = 1 - 1/60$ and hence it follows that
the scalar-tensor ratio is bounded from above as $r < 0.006$, which is very small.


\subsection{Comments on Initial Conditions}

As we have previously noted, our classical solution $\phi(t)$ sits at the unstable 
maximum of the potential for an infinite amount of time, thus this model would seem to admit 
infinitely many e-foldings of inflation.  Of course, this cannot be the case quantum mechanically
and one expects quantum fluctuations to displace $\phi$ from the false vacuum and cause it
to roll down the potential (we have assumed that this rolling takes place towards $\phi = 0$,
rather than down the unbounded side of the potential).  As a consistency check we note that
\be
  \frac{\langle (\delta \phi)^2 \rangle^{1/2}}{\phi_0} 
  = \langle (\delta \phi)^2 \rangle^{1/2} \sim \frac{H_0}{A}
\ee
We should compare this to $u_\star$ (see equation \ref{ustar}), the distance the field has
rolled classically at horizon crossing.  It is straightforward to show that
\[
  \frac{\langle (\delta \phi)^2 \rangle^{1/2}}{u_\star} \sim 
  10^{-5}\, |n_s-1|
\]
where we have used (\ref{cobe}).  Since $\langle (\delta \phi)^2 \rangle^{1/2} \ll u_\star$ 
for all parameter values it follows that 
the de Sitter space fluctuations (which are present as $u\rightarrow 0$) will not 
displace the field far enough from the maximum of the potential to have any significant
effect on the number of observable e-foldings of
inflation, although they prevent inflation from being
past-eternal.  We note, however, that if one incorporates thermal fluctuations
(and initial momentum) then
this model may suffer from problems related to fine tuning the intial conditions 
as in small field inflationary models \cite{small_field}.  However, it is not clear 
if these objections apply to our model for several reasons.  The first reason
is that the dynamics of this theory is peculiar and it is not clear how (or if) the  
phase space arguments of \cite{small_field} apply.  The second reason is that it is not 
clear what initial conditions for the field $\phi$ are most natural from a string theory 
perspective.  Finally we note that since the field $\varphi$ rolls a distance $A > M_p$ 
in field space, our model is not a ``small field'' model in the conventional 
sense.\footnote{The skeptical reader might have reservations
about the validity of our analysis since $A > M_p$.  We note that the action (\ref{action})
is not a low energy effective field theory and hence we believe that we are justified in using
this action even for super-Planckian symmetry breaking scale.}

If $p$-adic superstrings exist, it might also be possible to justify the initial conditions
for inflation by having topological inflation \cite{topological}, if the tachyon potential 
is symmetric about the unstable maximum.  This distinction exists between the tachyon of the 
open bosonic string \cite{bosonic} (describing the instability of D25 branes), and the tachyon 
of unstable branes in superstring theory \cite{susy}.  Any realistic extension of the model 
should have a potential which is bounded from below, and if it is supersymmetric, the minima 
should be at zero, hence the additional minima will be degenerate with the one at $\phi=0$.  
The existence of domains of the universe in the different minima ensures that there will be 
regions in between where inflation from the maximum of the potential is taking place, so long as 
the minima are discrete and not connected to each other by a continuous symmetry.

As we have noted previously the fact that the potential $V(\phi)$ is unbounded from below
is thought to be a reflection of the closed string tachyonic instability of bosonic
string theory.  If this conjecture is correct then the addition of supersymmetry should 
indeed lead to a symmetric potential for the $p$-adic tachyon which is bounded from below, as we 
have suggested above.

\section{Conclusions} 
\label{sec6} 
In this paper we have constructed for the first time approximate
solutions of the fully nonlocal $p$-adic string theory coupled to gravity, in which the
$p$-adic tachyon drives a sufficiently long period of inflation while rolling away from the
maximum of its potential.  In our solution, the nonlocal nature of the theory played an
essential role in obtaining slow-roll, since with a conventional kinetic term the potential
would have been too steep to give inflation.  One of the novel features of this construction
is that the Hubble parameter is larger than the string scale during inflation, a condition
which would usually invalidate an effective field theory description, but which
is consistent in the present context because of the ultraviolet-complete nature of the
theory.  

We found that the experimental constraints on the amplitude of the spectrum of
scalar perturbations produced by inflation require a small value of the string
coupling $g_s$, and can be consistent with a large range of values of the 
parameter $p$, $1\lsim p \lsim 10^{14}$.   The regime $p\gg 1$ is interesting because
it exhibits qualitatively different behavior relative to conventional inflationary 
models: slow roll despite the potential being steep, and inflation being driven
by the kinetic as well as potential energy of the field.  This regime is also interesting
because it corresponds to $g_s = \mathcal{O}(1)$ and hence appears more natural from
a string theory perspective.  Since the $p$-adic string is not construed
to be a realistic model by itself, it may not be very meaningful to question how
natural such values might be.  However, it may not be unreasonable to think
of real strings as being composed of constituent $p$-adic strings because the
Veneziano amplitude of the $p$-adic theory is related to that of the 
full bosonic string by ${\cal A}^{-1} = \prod_p
{\cal A}_p$ where the product is over all prime numbers $p$.   Thus it may 
not be unreasonable to expect similar behavior to the large-$p$ results from 
a more realistic model.

The model predicts a red spectrum, in agreement with the latest WMAP data,  whose
tilt is related to  the ratio of the string scale to the Hubble rate during
inflation via $H/m_s =  2/\sqrt{3|n_s-1|}$.  For $n_s=0.95$ this gives
$H/m_s \cong 5$.  This is in contrast to most stringy models of inflation which
require $H < m_s$ in order for the effective field theory to be valid.   We find
the bound $r < 0.006$ on the tensor modes.  It has been  estimated that future experiments 
could eventually have a  sensitivity of $r\sim 6\times 10^{-5}$ \cite{Cooray} and hence
the tensor components may in fact be observable.

We noted that the $p$-adic model succeeds with inflation where the real string theory
tachyon fails.  But our analysis makes it clear that this could be due to the failure to
keep terms with arbitrary numbers of derivatives in the action.  The effective tachyon
action of  Sen \cite{Sen:TachyonMatter} is a truncation which keeps arbitrary powers of first 
derivatives but ignores higher order derivatives, which were essential for obtaining our 
solution.  Thus the new features for inflation which we find in $p$-adic string theory could 
also be present in realistic string theories, if we knew how to include the whole tower of higher
dimensional kinetic terms.

\section*{Acknowledgments}

This work was supported in part by NSERC and FQRNT.   We would like to thank
R.\ Brandenberger, G.\ Calcagni, K.\ Dasgupta, A.\ Mazumdar and W.\ Siegel 
for interesting discussions, and D.\ Ghoshal for valuable comments on
the manuscript.

\section*{Note Added}

Upon completing this paper a related work appeared \cite{lidsey} in which an alternative
normalization for the fluctuations of the $p$-adic scalar was proposed.  
Motivated by this work, we reconsidered
our choice of normalization and concluded that (\ref{varphi},\ref{A}) is the most appropriate
definition of a ``canonical'' field.  Notice, however, that our field $\varphi$ differs from
the definition of a canonical field which was proposed in \cite{lidsey}.  
In \cite{lidsey} a field redefinition is advocated which puts the stress tensor $T_{\mu\nu}$ 
into canonical form, though this definition does not have canonical kinetic term in the 
action.  Though we believe that our definition is more natural, we stress that in the case
of current interest, $p$-adic inflation, the distinction does not generate any significant
\emph{quantitative} difference because our definition differs from that of \cite{lidsey} by a 
factor proportional to $\sqrt{\ln p}$ (which is less than an order of magnitude for the values
of $p$ which we consider).  We are grateful to J.\ Lidsey for sending us a draft
of his manuscript prior to publication and also for interesting and enlightening discussions.


\renewcommand{\theequation}{A-\arabic{equation}}
\setcounter{equation}{0}  

\appendix
\section{The Stress Energy Tensor and the Friedmann Equation}
\label{appa}

Here we compute the $O(u^2)$ term of the approximate solutions
\begin{eqnarray*}
  \phi &=& 1 - u - \phi_2 u^2 \label{phi_expand_app}\\
  H &=& H_0 - H_2 u^2 \label{H_expand_app}
\end{eqnarray*}
We write the different terms appearing in the energy density as 
$$
\rho_{\phi}=-T_{00}={m_s^4\over 2g_p^2}\LT \phi e^{-{\Box\over m_p^2}}\phi-{2\over p+1}\phi^{p+1}+{1\over m_p^2}\int_0^1 d\tau\ \LF \Box e^{-{\tau\Box\over m_p^2}}\phi\RF\LF e^{-{(1-\tau)\Box\over m_p^2}}\phi\RF\right.
$$
\be
\left.+{1\over m_p^2}\int_0^1 d\tau\ \p_t\LF e^{-{\tau\Box\over m_p^2}}\phi\RF\p_t\LF e^{-{(1-\tau)\Box\over m_p^2}}\phi\RF\RT\equiv {m_s^4\over 2g_p^2}(T_1+T_2+T_3+T_4)
\ee
where $T_1$, $T_2$, $T_3$ and $T_4$ are defined as in (\ref{T1}-\ref{T4}).  We now evaluate
this expression term-by-term.

We first consider $T_1 + T_2$, which can be written as
\[
  T_1 + T_2 = \frac{p-1}{p+1}\phi^{p+1}
\]
using (\ref{KG}).  Using (\ref{phi_expand_app}) we have
\begin{equation}
\label{T1+T2_app}
  T_1 + T_2 = \frac{p-1}{p+1} + (-p+1)u + (p-1)\left[-\phi_2 + \frac{p}{2}\right]u^2 
  + \mathcal{O}(u^3)
\end{equation}

To evaluate $T_3$, $T_4$ we note that
\begin{eqnarray}
  e^{-\tau\Box/m_p^2}\phi &=& 1-e^{\tau\mu_1}u - \phi_2 e^{\tau\mu_2} u^2 + \cdots \nonumber \\
  e^{-(1-\tau)\Box/m_p^2}\phi &=& 1-e^{(1-\tau)\mu_1}u - \phi_2 e^{(1-\tau)\tau\mu_2} u^2 + \cdots
  \label{tau_app}
\end{eqnarray}
(see equation \ref{eboxphi} with $\sigma = 0$).  It is straightforward to show that
\begin{equation}
\label{box_tau_app}
  \Box e^{-\tau\Box /m_p^2}\phi = e^{\tau\mu_1}(\lambda^2 + 3H_0\lambda) u +
                                  \phi_2 e^{\tau\mu_2}(4\lambda^2 + 3H_0\lambda) u^2 + \cdots
\end{equation}

We are now in a position to compute $T_3$.  Using (\ref{tau_app}) and (\ref{box_tau_app})
we have
\begin{eqnarray*}
  T_3 &=& \frac{1}{m_p^2} \int_0^1 d\tau \left[  
          (\lambda^2 + 3H_0\lambda) e^{\tau\mu_1} u \right. \\
          &+& \left. \left[-(\lambda^2 + 3H_0\lambda)e^{\mu_1} 
          + \phi_2(4\lambda^2 + 6H_0\lambda)e^{\tau\mu_2}\right]u^2 + \cdots \right]
\end{eqnarray*}
The $d\tau$ integrals are trivial to perform using the identity
\begin{equation}
  \int_0^1d\tau e^{\alpha\tau} = \frac{e^{\alpha}-1}{\alpha}
\end{equation}
We find that
\begin{equation}
\label{T3_app}
  T_3 = (p-1) u + \left[-p\mu_1 + (e^{\mu_2}-1)\phi_2\right] u^2 + \mathcal{O}(u^3)
\end{equation}

We now consider the integrand of $T_4$:
\be
\p_t\LF e^{-{\tau\Box\over m_p^2}}\phi\RF\p_t\LF e^{-{(1-\tau)\Box\over m_p^2}}\phi\RF\cong \la^2 u^2\p_u\LF1-e^{\tau\mu_1}u\RF\p_u\LF1-e^{(1-\tau)\mu_1}u \RF=\la^2 u^2e^{\mu_1}
\ee
The $d\tau$ integral is trivial and gives 
\be
\label{T4_app}
T_4= u^2{\la^2\over m_p^2}e^{\mu_1}
\ee

It is straightforward to sum up the various contributions to $T_{00}$.  We find that
\begin{eqnarray}
  \rho_\phi &=& \frac{m_s^4}{2 g_p^2} 
                \left[ \frac{p-1}{p+1} \right. \nonumber \\
            &+& \left. \left[ (-p+e^{\mu_2})\phi_2 + \frac{p}{2}(p-1) - p\ln p 
                 + p \frac{\lambda^2}{m_p^2}\right] u^2 + \cdots \right] \label{rho_u_app}
\end{eqnarray}
The fact that the coefficient of the $\mathcal{O}(u)$ term is zero verifies that $H_1 = 0$.
We now solve the Friedmann equation
\[
  3 H^2 = \frac{1}{M_p^2}\rho_\phi
\]
noting that
\[
  H^2 = H_0^2 - 2H_0 H_2 u^2 + \mathcal{O}(u^3)
\]
Matching the coefficients at order $u^0$ gives
\[
  H_0^2 = \frac{1}{6g_p^2}\frac{m_s^4}{M_p^2}\frac{p-1}{p+1}
\]
as before.  Matching the coefficients at order $u^2$ gives
\begin{eqnarray*}
  -6H_0H_2 &=& \frac{m_s^4}{2g_p^2 M_p^2}
               \left[ (e^{\mu_2}-p)\phi_2 + \frac{p}{2}(p-1) 
           - p\ln p + p \frac{\lambda^2}{m_p^2} \right] \\
           &=& \frac{m_s^4}{M_p^2}\frac{p}{2g_p^2}\left[
               \frac{\lambda^2}{m_p^2} - \ln p \right] \\
           &=& \frac{m_s^4}{M_p^2}\frac{p}{2g_p^2}\left[ \frac{-3H_0\lambda}{m_p^2} \right] 
\end{eqnarray*}
In the first equality we have used that fact that
\[
  \phi_2 e^{\mu_2} = p\phi_2 - \frac{p}{2}(p-1)
\]
which follows from the second order Klein-Gordon equation and in the second equality we
have used
\[
  \frac{\lambda^2}{m_p^2} + 3\frac{H_0\lambda}{m_p^2} = \mu_1 = \ln p
\]
which follows from the first order Klein-Gordon equation.
Finally we arive at the result for $H_2$:
\begin{equation}
  H_2 = \frac{m_s^4}{4g_p^2M_p^2}\frac{p\lambda}{m_p^2} 
  = \frac{1}{8 g_s^2}\frac{p^3\ln p}{p-1}\left(\frac{m_s}{M_p}\right)^2\lambda
\end{equation}

\renewcommand{\theequation}{B-\arabic{equation}}
\setcounter{equation}{0}  

\section{Incomplete Cylindrical Functions of the Sonine-Schaefli Form}
\label{appb}

The Bessel function can be represented by a contour integral of the Sonine-Schaefli form
\begin{equation}
\label{Jnu}
  J_\nu(z) = \frac{1}{2\pi i}\left(\frac{z}{2}\right)^\nu 
  \int_{c-i\infty}^{c+i\infty} \omega^{-\nu - 1} 
  \exp\left[ \omega - \frac{z^2}{4\omega}\right]d\omega
\end{equation}
as long as $\mathrm{Re}(\nu) > -1$.  In (\ref{Jnu}) $c$ is an arbitrary positive constant.
The incomplete cylindrical function, $S_\nu(r,s;z)$ generalizes (\ref{Jnu}) to arbitrary limits 
of integration
\begin{equation}
\label{Snu}
  S_\nu(r,s;z) = \frac{1}{2\pi i}\left(\frac{z}{2}\right)^\nu 
  \int_{r}^{s} \omega^{-\nu - 1} \exp\left[\omega - \frac{z^2}{4\omega}\right]d\omega
\end{equation}
It follows that
\begin{equation}
\label{integral}
  \int_{r}^{s} \omega^{-\nu - 1} \exp\left[-\omega - \frac{z^2}{4\omega}\right]d\omega
   = 2\pi i e^{i\nu\pi/2}\left(\frac{z}{2}\right)^{-\nu} S(-r,-s;iz)
\end{equation}
Taking the limits $r\rightarrow 0$ and $s \rightarrow +\infty$ this integral can be written
in terms of Hankel functions of imaginary argument, $K_\nu(z)$, as
\begin{equation}
\label{hankel}
  \int_{0}^{\infty} \omega^{-\nu - 1} \exp\left[-\omega - \frac{z^2}{4\omega}\right]d\omega
   = 2\left(\frac{z}{2}\right)^{-\nu} K_\nu(z)
\end{equation}
The function $K_\nu(z)$ is real-valued for real $z$ and is related to the usual Hankel 
function as $K_\nu(z) = (i\pi / 2) e^{i\pi\nu/2} H_\nu^{(1)}(iz)$.  Using the well known 
large-argument asymptotics of the Hankel functions one may show that
\begin{equation}
\label{large_z}
  \int_{0}^{\infty} \omega^{-\nu - 1} \exp\left[-\omega - \frac{z^2}{4\omega}\right]d\omega
  \cong \sqrt{\frac{2\pi}{z}}\left(\frac{z}{2}\right)^{-\nu} e^{-z}
\end{equation}
for $z \gg 1$.

\renewcommand{\theequation}{C-\arabic{equation}}
\setcounter{equation}{0}  

\section{Comparison to Local Field Theory}
\label{appc}

In this appendix we perform a detailed comparison of our results for the action (\ref{action})
to the theory
\begin{equation}
\label{chi4}
  S = - \int d^4 x \left[\frac{1}{2}\partial_\mu\psi\partial^\mu\psi + V(\psi) \right]
\end{equation}
with potential
\begin{eqnarray}
  V(\psi) &=& \frac{g_s}{4}\left(\psi^2 - v^2 \right)^2 \nonumber \\
          &\equiv& \frac{g v^4 }{4} - m_s^2\psi^2 + \frac{g}{4}\psi^4
  \label{pot_app}
\end{eqnarray}
where we have defined $m_s^2 \equiv 2 g_s v^2$.  We are interested in obtaining inflation near the
unstable maximum $\psi = 0$.  The flatness of the potential is parameterized by the 
dimensionless slow roll parameters
\begin{eqnarray}
  \epsilon &=& \frac{M_p^2}{2}\left(\frac{V''(\psi)}{V(\psi)}\right)^2 = 
  \frac{8\psi^2M_p^2}{(\psi^2-v^2)^2} \label{eps_app}\\
  \eta &=& M_p^2\frac{V''(\psi)}{V(\psi)} 
  = \frac{4 M_p^2}{(\psi^2-v^2)^2}\left[3\psi^2-v^2\right]
  \label{eta_app}
\end{eqnarray}
Unlike in the $p$-adic theory we do not need to distinguish between $\epsilon_\sH,\eta_\sH$ and 
$\epsilon_V,\eta_V$ (see equations \ref{epsilon}-\ref{eta_V}) and hence we drop the subscripts
on the slow roll parameters.
For $\psi \ll v$ the $\epsilon$ parameter (\ref{eps_app}) is automatically small while
\[
  |\eta| \cong 8\left(\frac{M_p}{v}\right)^2
\]
which is small compared to unity for $v \gg M_p$.  The fact that the symmetry breaking scale
is large compared to the Planck scale may be reason to doubt the validity of the field theory 
(\ref{chi4}).  However, for our purposes this is irrelevant.

It is instructive to consider solving the equations of motion for this theory using the formalism
of section \ref{solns}.  We begin by speculating solutions of the form
\begin{eqnarray}
  \psi &=& v e^{\lambda t} + \psi_2 v e^{2\lambda t} + \psi_3 v e^{3\lambda t}
           + \mathcal{O}(e^{3\lambda t}) \label{chi_series}\\
  H &=& H_0 - H_1 e^{\lambda t} - H_2 e^{2\lambda t} + \mathcal{O}(e^{3\lambda t}) 
\label{H_series_app}
\end{eqnarray}
and solving order by order in $e^{\lambda t}$.  
The ansatz (\ref{chi_series},\ref{H_series_app}) is analogous to (\ref{phi_series},\ref{H_series})
since in both cases the (classical) field spends an infinite amount of time at the unstable
maximum, driving a past-eternal de Sitter phase, before rolling towards the true minimum of
the potential.
We suppose that $e^{\lambda t} \ll 1$ so that $\psi \ll v$ initially.

The Klein-Gordon equation is
\[
  \frac{\ddot{\psi}}{v} + 3 H \frac{\dot{\psi}}{v} =  2  m_s^2 \frac{\psi}{v} 
  + 2 m_s^2 \left(\frac{\psi}{v}\right)^3
\]
Plugging in the ansatz (\ref{chi_series}) we obtain
\begin{equation}
\label{l_app}
  \lambda^2 + 3 H_0 \lambda = 2m_s^2
\end{equation}
at first order in $e^{\lambda t}$.  This result is identical to (\ref{la_simple}).  
At second and third order respectively we find
\begin{eqnarray*}
  \psi_2 &=& 0 \\
  \psi_3 &=& \frac{2 m_s^2 + 3 H_2 \lambda}{9\lambda^2 - 2 m_s^2}
\end{eqnarray*}

The Friedmann equation is
\[
  3 H^2 = \frac{1}{M_p^2}\left[\frac{g v^4}{4} - m_s^2 v^2 \left(\frac{\psi}{v}\right)^2 
  + \frac{g v^4}{4}\left(\frac{\psi}{v}\right)^4 
  + \frac{v^2}{2}\left(\frac{\dot{\psi}}{v}\right)^2\right]
\]
At zeroth order in $e^{\lambda t}$ we obtain the familiar result
\begin{equation}
\label{H0_app}
  H_0 = \frac{g^{1/2}}{2\sqrt{3}}\frac{v^2}{M_p}
\end{equation}
At first and second order we find that 
\begin{eqnarray}
  H_1 &=& 0 \label{H1_app}\\
  H_2 &=& \frac{2}{\sqrt{3 g}}\frac{2 m_s^2-\lambda^2}{M_p} 
          = \frac{1}{4}\left(\frac{v}{M_p}\right)^2 \lambda\label{H2_app}
\end{eqnarray}
In writing the second equality in (\ref{H2_app}) we have used 
$\lambda^2 + 3H_0 \lambda = 2 m_s^2$.

Though a complete treatment of inhomogeneities including metric perturbations and
nonzero slow roll parameters is straightforward in this context we choose to analyze 
this theory using the same approximations as we used in section \ref{fluct} in order
to make more explicit the comparison between the two theories.
The perturbed Klein-Gordon equation (neglecting metric fluctuations and in the limit
$e^{\lambda t} \rightarrow 0$) is
\[
  \left(\partial_t^2 + 3H_0\partial_t - \frac{\partial_i\partial^i}{a^2} - 2 m_s^2\right)  
  \delta\psi = 0
\]
where $a = e^{H_0 t}$.  The large scale solution is
\[
  |\delta\psi_k| \cong \frac{H}{\sqrt{2k^3}}\left(\frac{k}{a H_0}\right)^{3/2 - \nu}
\]
where
\[
  \nu = \sqrt{\frac{9}{4} + \frac{2 m_s^2}{H_0^2}}
\]
Near scale-invariance of the spectrum requires $m_s \ll H_0$.  In this limit we have
\begin{equation}
\label{n_appendix}
  n_s-1 \cong -\frac{4}{3}\frac{m_s^2}{H_0^2} = -8\left(\frac{M_p}{v}\right)^2
\end{equation}
which is identical to (\ref{n}).  This result also reproduces the full calculation incorporating
metric perturbations: $n_s - 1 \cong 2\eta - 6\epsilon \cong -8M_p^2 / v^2$.

We can use (\ref{n_appendix}) to write the dimensionless quantities $H_0/m_s$, $H_2/m_s$ and
$\lambda / m_s$ in terms of $|n_s-1|$.  The solution of (\ref{l_app}) can be written as
\[
  \frac{\lambda}{m_s} \cong \sqrt{\frac{|n_s-1|}{3}}
\]
which is identical to (\ref{lambda}).  For this solution we of course have 
$3\lambda H_0 \gg \lambda^2$ so that the evolution is friction dominated.
It is also straightforward to show that
\begin{eqnarray*}
  \frac{H_0}{m_s} &=& \frac{2}{\sqrt{3 |n_s-1|}} \\
  H_2 &=& H_0
\end{eqnarray*}
We see that the solutions $H$, $\chi$ for the theory (\ref{chi4}) are identical to those
of the theory (\ref{action}) up to order $e^{\lambda t}$.  At order $e^{2\lambda t}$ and higher,
however, the dynamics of the two theories differs.

From (\ref{eps_app}) one can check that $\epsilon \cong 1$ at $\psi \lsim v$ so that it
is a good approximation to suppose that inflation ends at $e^{\lambda t} \cong 1$.  
It is straightforward to impose the COBE normalization $V / (\epsilon M_p^4) = 6\times 10^{-7}$
for this model (which will impose $g_s \ll |n_s-1|^2$), however, this is unnecessary for our 
purposes.  It is clear that the inflationary dynamics and predictions predictions of the 
theory (\ref{chi4}) are identical to those of the theory (\ref{action}).



\bibliographystyle{apsrmp}
\bibliography{rmp-sample}

\end{document}